\documentclass[aps,pra,twocolumn,superscriptaddress,showpacs,nofootinbib,amsmath,amssymb]{revtex4}
\usepackage{mathrsfs}

\newcommand{\h}{\hat{H}}
\newcommand{\Hcl}{\mathcal{H}}
\newcommand{\Pcl}{\boldsymbol{\mathcal{P}}}

\renewcommand{\P}{\mathbf{p}}
\newcommand{\R}{\mathbf{r}}
\newcommand{\K}{\mathbf{k}}
\newcommand{\A}{\mathbf{A}}

\newcommand*{\bra}[1]{\left\langle{#1}\right|}
\newcommand*{\ket}[1]{\left| #1 \right\rangle}

\renewcommand{\Im}{{\rm Im}\,}

\begin{document}
 
\title{Adaptation of the Modified Adiabatic Approximation to Strong Field Ionization}

\author{Denys I. Bondar}
\email{dbondar@sciborg.uwaterloo.ca}
\affiliation{University of Waterloo, Waterloo, Ontario N2L 3G1,
Canada}
\affiliation{National Research Council of Canada, Ottawa,
Ontario K1A 0R6, Canada}

\author{Wing-Ki Liu}
\email{wkliu@sciborg.uwaterloo.ca}
\affiliation{University of Waterloo, Waterloo, Ontario N2L 3G1,
Canada}

\author{Gennady L. Yudin}
\email{gennady.yudin@nrc.ca}
\affiliation{National Research Council of Canada, Ottawa, Ontario K1A 0R6, Canada}
\affiliation{Universit\'{e} de Sherbrooke, Sherbrooke, Qu\'{e}bec J1K 2R1, Canada}

\date{\today}

\begin{abstract}
Savichev's modified adiabatic approximation [Sov. Phys. JETP {\bf 73}, 803 (1991)] is used to obtain a general form of a quantum-mechanical amplitude of ionization by a low-frequency laser field. The method possesses only one requirement that the frequency of the laser field must be low. Connections of the obtained result with the quasi-classical approximation and other previous investigations are discussed.
\end{abstract}

\pacs{32.80.Rm, 32.80.Fb}

\maketitle


Quantum-mechanical phenomena induced in a strong laser field are explicitly non-perturbative in nature \cite{Krausz2009}. Needless to say, theoretical investigations of such strong field phenomena go beyond the conventional perturbation theory. Beside numerical integration of the Schr\"{o}dinger equation, many varieties of the quasi-classical  (QC) approximation are widely employed \cite{Krausz2009, Popov2004, Popov2005}. In the current paper, we focus our attention on another approach -- the adiabatic approximation.

Roughly speaking, the adiabatic approximation can be introduced as follows. Once the frequency of the external laser field is much lower than the characteristic atomic frequency, $\omega \ll \omega_{at}$, an approximate solution of the Schr\"{o}dinger equation can be found by means of averaging over atomic (internal) degrees of freedom. Therefore, the adiabatic approximation is a method of constructing the asymptotic expansion of the solution of the non-stationary Schr\"{o}dinger equation in terms of the small parameter $\omega/\omega_{at}$.

Born and Fock \cite{Born1928} founded the theory of the adiabatic approximation for a discrete spectrum by formulating the adiabatic theorem. Landau \cite{Landau1932} estimated the probability of nonadiabatic transitions between discreet states. However, the leading-order asymptotic result for such a quantity was obtained by Dykhne \cite{Dykhne_1962}. Afterwards,  nonadiabatic transitions between discreet states was thoroughly analyzed by many authors \cite{Davis_1976, Solovev1976a} (for reviews see Ref. \cite{ENikitin1984}). Summarizing research in this area, one may conclude that nonadiabatic transitions in discreet spectra are quite well studied.

Nevertheless, transitions from a discreet state to a continuum one are the subject of on-going investigations for many decades. Despite much work in this topic, a universally accepted approach is lacking. The direct generalization of the Dykhne method was developed by Chaplik \cite{Chaplik_1964}. Exactly solvable models were reported by Demkov and Oserov \cite{Demkov1966},  
Ostrovskii \cite{Ostrovskii1972}, and Nikitin \cite{Nikitin1962} (for review see, e.g., Ref. \cite{Nikitin1970}). The advanced adiabatic approach was introduced and widely employed by Solov'ev \cite{Solovev1976}. Finally, Tolstikhin recently developed a promising version of the adiabatic approximation for the transitions to the continuum \cite{Tolstikhin2008a}  by employing the Siegert-state expansion for nonstationary quantum systems \cite{Tolstikhin2006}. Yet, the approach presented in Ref. \cite{Tolstikhin2008a} is limited to finite range potentials. 


Our investigations are based on a seminal result that ought to be summarized foremost. Following the Solov'ev advanced adiabatic approach \cite{Solovev1976}, Savichev \cite{Savichev1991a} proved the following. If the adiabatic state $\ket{\psi_i(\varphi)}$ and the corresponding adiabatic term $E_i(\varphi)$,
 \begin{eqnarray}\label{AdiabaticTermStateFullH}
 \h(\varphi)\ket{\psi_i (\varphi)} = E_i(\varphi)\ket{\psi_i(\varphi)},
 \end{eqnarray}
are known, then the solution $\ket{\Psi(t)}$ of the nonstationary Schr\"{o}dinger equation  (the atomic units, $\hbar=m=|e|=1$, are used throughout unless stated otherwise)
\begin{equation}\label{FullSchEq}
 i\partial_t \ket{\Psi(t)} = \h(\varphi)\ket{\Psi(t)},
 \end{equation}
 where $\varphi = \omega t$ is a phase of the laser field,
subjected to the initial condition
\begin{eqnarray}\nonumber
\ket{\Psi(t)} \xrightarrow[t\to -\infty]{} \ket{\psi_i(\varphi_i(E_i))}e^{ - i\int^t E_i(\omega\tau)d\tau} [ 1+ O(\omega)],
\end{eqnarray}
has the following form within the adiabatic approximation ($\omega \ll 1$)
\begin{eqnarray}\label{SavichevEq}
&& \ket{\Psi(t)} = \frac 1{2\pi\omega}\iint dEd\varphi'  \, \ket{\psi_i (\varphi_i(E))}[ 1 + O(\omega)]\times \nonumber\\
&& \qquad\exp\left[ \frac i{\omega}\left(E\varphi' - \int^{\varphi'}E_i(\varphi)d\varphi \right)-iEt\right].
\end{eqnarray}
Here $\varphi_i(E)$ is the inverse function of $E_i(\varphi)$. Note that no assumptions on a form of the Hamiltonian $\h$ were made.

Let $\ket{i}$ and $\ket{f}$ be stationary states (for specification see Eq. (\ref{IniConditions}) and the comment after), and we shall assume that the quantum system with the Hamiltonian $\h$ is in the state $\ket{i}$ at $t=-\infty$. The main aim of this section is to obtain the general form of the transition amplitude $\mathfrak{M}_{i\to f}$ that the given quantum system will be found in the state $\ket{f}$ at $t=+\infty$.

Before going further, we are to introduce notations. First, we arbitrarily partition the Hamiltonian $\h$:
\begin{equation}\label{Partition}
\h(\varphi) \equiv \h_0(\varphi) + \hat{V}(\varphi).
\end{equation}
Second, we denote by $\ket{\Psi_{i,f}(t)}$  the solutions of Eq. (\ref{FullSchEq}) such that 
$\ket{\Psi_i(-\infty)} = \ket{i}$, $\ket{\Psi_f(+\infty)} = \ket{f}$; similarly, $\ket{\Phi_{i,f}(t)}$ are the solutions of  the nonstationary Schr\"{o}dinger equation
$$
 i\partial_t \ket{\Phi(t)} = \h_0(\varphi)\ket{\Phi(t)},
 $$
 with the initial conditions: $\ket{\Phi_i(-\infty)} = \ket{i}$ and $\ket{\Phi_f(+\infty)} = \ket{f}$, correspondingly.
 
Having defined all necessary functions, we introduce two equivalent forms of the transition amplitude $\mathfrak{M}_{i\to f}$ by employing the corresponding version of the $S$-matrix (see, e.g., Ref. \cite{Becker2005a}):
the reversed time form (sometimes called the ``prior'' form) 
\begin{equation}\label{M_PriorForm}
\mathfrak{M}_{i\to f}^{(r)} = -i \int_{-\infty}^{\infty}  \bra{ \Psi_f (t)}\hat{V}(\omega t)\ket{\Phi_i (t)}dt 
\end{equation}
and the direct time form (the ``post'' form)
\begin{equation}\label{M_PostForm}
\mathfrak{M}_{i\to f}^{(d)} = -i \int_{-\infty}^{\infty}  \bra{ \Phi_f (t)}\hat{V}(\omega t)\ket{\Psi_i (t)}dt.
\end{equation}
It is noteworthy to recall the physical interpretation of Eqs. (\ref{M_PriorForm}) and (\ref{M_PostForm}). The terms $\bra{ \Psi_f (t)}\hat{V}(\omega t)\ket{\Phi_i (t)}$ and $\bra{ \Phi_f (t)}\hat{V}(\omega t)\ket{\Psi_i (t)}$ can be regarded as the amplitudes of quantum ``jumps,'' which occur at the time moment $t$. The integrals over $t$ convey that these jumps take place at {\it any} time.

Introducing the adiabatic state $\ket{\phi_f(\varphi)}$ and term $E_f(\varphi)$ of the Hamiltonian $\h_0$,
\begin{equation}\label{H0AdiabaticState}
\h_0(\varphi) \ket{\phi_f(\varphi)} = E_f (\varphi) \ket{\phi_f(\varphi)}, 
\end{equation}
the wave function $\ket{\Phi_f(t)}$ can be readily presented in the form of Eq. (\ref{SavichevEq}). In further investigations, we employ the post form [Eq. (\ref{M_PostForm})], and thus we shall assume that 
\begin{equation}\label{IniConditions}
\ket{\psi_i(-\infty)} \equiv \ket{i}, \quad \ket{\phi_f(+\infty)} \equiv \ket{f}.
\end{equation}
In the case of the prior form [Eq. (\ref{M_PriorForm})], condition (\ref{IniConditions}) has to be substituted by 
$\ket{\phi_i(-\infty)} \equiv \ket{i}$ and $\quad \ket{\psi_f(+\infty)} \equiv \ket{f}$, where $\ket{\phi_i(\varphi)}$ and $\ket{\psi_f(\varphi)}$ are adiabatic states of the Hamiltonians $\h_0$ and $\h$, correspondingly. 

Substituting the asymptotic representations [Eq. (\ref{SavichevEq})] of the wave functions $\ket{ \Phi_f (t)}$ and $\ket{\Psi_i (t)}$ into Eq. (\ref{M_PostForm}), we obtain 
\begin{eqnarray}\label{M_Before_SaddlePointInt}
\mathfrak{M}_{i\to f}^{(d)} = \frac{-i}{(2\pi)^2\omega^3} 
\int f({\bf z}) e^{iS({\bf z})/\omega} d^5 {\bf z}\left[ 1+ O(\omega)\right],
\end{eqnarray}
where ${\bf z} = (\mathcal{E}, \eta, \mathcal{E}', \eta', \varphi)$ is a five-dimensional vector, $d^5 {\bf z} = d\mathcal{E} d\eta d\mathcal{E}' d\eta' d\varphi$, 
$f({\bf z}) = \bra{\phi_f\left(\varphi_f(\mathcal{E}')\right)}\hat{V}(\varphi)\ket{\psi_i\left(\varphi_i(\mathcal{E})\right)}$, and
$S({\bf z}) = \mathcal{E}\left(\eta-\varphi\right) + \mathcal{E}'\left(\varphi-\eta'\right) - \int^{\eta} E_i(\xi)d\xi + \int^{\eta'} E_f(\xi)d\xi$. Bearing in mind that $1/\omega$ is a large parameter, the five-dimensional integral in Eq. (\ref{M_Before_SaddlePointInt}) can be calculated by means of the saddle-point approximation. Finally,   {\it the post form of the transition amplitude} within the adiabatic approximation reads
 \begin{eqnarray}\label{AmplitudeFinalExpr}
\mathfrak{M}_{i\to f}^{(d)} = \sqrt{\frac{2\pi}{\omega}}\sum_{\varphi_{\star}} 
\frac{ \bra{\phi_f(\varphi_{\star})} \hat{V} (\varphi_{\star}) \ket{\psi_i(\varphi_{\star})}}
{  \left. \sqrt{ \frac{d}{d\varphi} \left[ E_f(\varphi)-E_i(\varphi)\right] } \right|_{\varphi = \varphi_{\star}} } \nonumber\\
\times\exp\left\{\frac i{\omega}\int^{\varphi_{\star}} \left[ E_f(\varphi) - E_i(\varphi) \right]d\varphi \right\} [ 1 + O(\omega)],
 \end{eqnarray}
 where $\sum_{\varphi_{\star}}$ denotes the summation over simple saddle points $\varphi_{\star}$, i.e., solutions of the equation 
 \begin{eqnarray}\label{SaddlePointEq}
 && E_f(\varphi_{\star}) = E_i(\varphi_{\star}), \\
 && \frac{d}{d\varphi} E_f(\varphi_{\star}) \neq \frac{d}{d\varphi}E_i(\varphi_{\star}). \label{SimpleSaddlePointEq}
 \end{eqnarray} 
 
The physical interpretation of the sum over $\varphi_{\star}$ is as follows: quantum jumps occur only at {\it isolated} time moments  $t_{\star}= \varphi_{\star}/\omega$, when the jumps are most probable; hence, $t_{\star}$ are called ``transition times.'' Note that the given interpretation deviates from the physical meaning of the time integral in Eq. (\ref{M_PostForm}).
 
 Some general remarks on Eq. (\ref{AmplitudeFinalExpr}), the main result of this paper, are to be made:

(i) $\varphi_{\star}$ is usually a complex solution of Eq. (\ref{SaddlePointEq}); therefore, saddle points $\varphi_{\star}$ with negative imaginary parts should be ignored because such points make exponentially large contributions to the amplitude, which leads to unphysical probabilities.

(ii) If $\varphi_{\star}^1, \, \varphi_{\star}^2, \dots, \varphi_{\star}^n$ are solutions of Eq. (\ref{SaddlePointEq}), such that 
$\Im\left(\varphi_{\star}^1\right)>\Im\left(\varphi_{\star}^2\right)> \ldots > \Im\left(\varphi_{\star}^n\right)>0$, then all but the single term that corresponds to the saddle point $\varphi_{\star}^n$ may be neglected in the sum over $\varphi_{\star}$ in Eq. (\ref{AmplitudeFinalExpr}). One is eligible to do so since this saddle point has the largest contribution to the transition  amplitude.

(iii) On the one hand, the explicit form of $E_f(\varphi)$ is solely determined by partitioning [Eq. (\ref{Partition})]; on the other hand, $E_i(\varphi)$ is unique for a given quantum system.

(iv) The exponential factor of Eq. (\ref{AmplitudeFinalExpr}) is similar to the exponential factor in the Dykhne approach  \cite{Dykhne_1962, Davis_1976, Chaplik_1964} (see also Refs. \cite{Landau_1977, Delone_1985}) -- the methods for calculating the amplitude of bound-bound transitions within the adiabatic approximation. Hence, Eq. (\ref{AmplitudeFinalExpr}) may be considered as a generalization of the Dykhne formula for bound-free transitions.

(v)  By employing an appropriate version of the saddle-point method, one can in principle generalize Eq. (\ref{AmplitudeFinalExpr}) for the case when condition (\ref{SimpleSaddlePointEq}) is violated.


Now, the connection between the amplitude [Eq. (\ref{AmplitudeFinalExpr})] and the method of complex trajectories is to be manifested. According to the method of complex classical trajectories (see, e.g., Refs. \cite{Landau1932, Nikitin1993, Landau_1977}, and the imaginary time method \cite{Popov2005}), to calculate the probability of the transition from the initial state to the final, one should first solve the corresponding  classical equations of motion and find the ``path'' of such a transition. However, this path is complex; in particular, the transition point $\R_{\star}$ and transition time $t_{\star}$ at which the transition occurs are complex. Parameters $\R_{\star}$ and $t_{\star}$ are determined by the classical conservation laws. Next, one has to obtain the classical action $S_f(\R_f, t_f; \R_{\star}, t_{\star}) + S_i(\R_{\star}, t_{\star}; \R_i, t_i)$ for the motion of the system in the initial state from the initial position $\R_i$ at time $t_i$ to the transition point $\R_{\star}$ at time $t_{\star}$ and then in the final state from $\R_{\star}$ at $t_{\star}$ to the final position $\R_f$ at time $t_f$. Finally, the probability of the transition is given by 
\begin{equation}\label{ImTime}
\Gamma \propto \exp\left\{ -2\Im\left[S_f(\R_f, t_f; \R_{\star}, t_{\star}) + S_i(\R_{\star}, t_{\star}; \R_i, t_i)\right]\right\}.
\end{equation}
Equations (\ref{AmplitudeFinalExpr}) and (\ref{ImTime}) must coincide in some region of parameters. The method of complex trajectories can be derived as the QC approximation of the transition amplitude [Eq. (\ref{M_PriorForm}) or Eq. (\ref{M_PostForm})]; we outline this derivation below. Therefore, it would be of methodological interest to establish an explicit connection between Eqs. (\ref{AmplitudeFinalExpr}) and (\ref{ImTime}). This connection has not been demonstrated. 

Without loss of generality, assuming that $\hat{V}(\omega t)$ is a non-differential operator, we obtain the QC approximation to Eq. (\ref{M_PostForm}) 
\begin{eqnarray}\label{M_WKB}
&& \mathfrak{M}_{i\to f}^{(d)} \approx -i\int dt \int d^3 \R d^3 \R_f d^3 \R_i \,\bra{f} \R_f \rangle \hat{V}(\omega t, \R) \bra{\R_i} i \rangle \nonumber\\
&& \quad\times F_f^* F_i \exp\left\{ i\left[S_f(\R_f, t_f; \R, t) + S_i(\R, t; \R_i, t_i) \right]\right\},
\end{eqnarray}
where $t_{f,i} = \pm\infty$, $F_f\exp(iS_f)$, and $F_i\exp(iS_i)$ are the QC versions of the propagators with the Hamiltonian $\h_0$ and $\h$, correspondingly.  We recall that the general form of the QC propagator is given by
\begin{eqnarray}\label{GenaralQCPropagator}
\sum_{\alpha} F^{(\alpha)}(\R, t; \R', t')\exp\left[iS^{(\alpha)}(\R, t; \R', t')\right],
\end{eqnarray}
where the sum denotes the summation over classical paths that connect the initial $(\R', t')$ and final $(\R,t)$ points. Therefore, usage of this form of the QC propagator, $F\exp(iS)$, is justified if we assume that there is only one such path; indeed, this is the case in the majority of practical calculations, and thus we shall accept this assumption hereinafter.

In order to reach Eq. (\ref{ImTime}) from  Eq. (\ref{M_WKB}), one has to calculate the integrals over $\R$ and $t$ in Eq. (\ref{M_WKB}) by means of the saddle-point approximation. The equations for the saddle points $\R_{\star}$ and $t_{\star}$, i.e., the transition points, read
\begin{eqnarray}
\left. \partial_t \left[ S_i(\R, t; \R_i, t_i) -S_f(\R, t; \R_f, t_f)  \right]\right|_{t=t_{\star},\, \R=\R_{\star}} = 0, \label{Saddle_PointT}\\
\left. \nabla_{\R}\left[ S_i(\R, t; \R_i, t_i) -S_f(\R, t; \R_f, t_f)  \right]\right|_{t=t_{\star},\, \R=\R_{\star}} = {\bf 0}. \label{Saddle_PointR}
\end{eqnarray}
Recalling the Hamilton-Jacobi equation
\begin{equation}\label{HJEq}
\partial_t S_{i,f} (\R, t; \R_{i,f}, t_{i,f}) = 
-\Hcl_{i,f}(\R,\Pcl_{i,f}, t), 
\end{equation}
where $\Hcl_{i,f}$ are classical Hamiltonians and $\Pcl_{i,f}$ are classical canonical momenta
$$
\Pcl_{i,f}(\R, t)  = \nabla_{\R} S_{i,f}(\R, t; \R_{i,f}, t_{i,f}),
$$
we rewrite Eqs. (\ref{Saddle_PointT}) and (\ref{Saddle_PointR}) as the law of conservation of canonical momentum and the law of conservation of energy:
\begin{eqnarray}
&& \Pcl_f(\R_{\star}, t_{\star}) = \Pcl_i(\R_{\star}, t_{\star}), \\
&& \Hcl_{f}(\R_{\star},\Pcl_{f}(\R_{\star}, t_{\star}), t_{\star}) = \Hcl_{i}(\R_{\star},\Pcl_{i}(\R_{\star}, t_{\star}), t_{\star}).
\end{eqnarray}

Having introduced all the necessary quantities, we demonstrate the correspondence between Eqs. (\ref{AmplitudeFinalExpr}) and (\ref{ImTime}) within exponential accuracy. Performing a simple transformation and using Eq. (\ref{HJEq}), we reach
\begin{eqnarray}
&& S_f(\R_f, t_f; \R_{\star}, t_{\star}) + S_i(\R_{\star}, t_{\star}; \R_i, t_i) \nonumber\\
&&=S_f(\R_f, t_f; \R_{\star}, t_i)+S_i(\R_{\star}, t_i; \R_i, t_i)  \nonumber\\
&&+ \int_{t_i}^{t_{\star}} \left[ \partial_{\tau} S_i(\R_{\star}, \tau; \R_i, t_i) -
\partial_{\tau} S_f(\R_{\star}, \tau; \R_f, t_f) \right]d\tau  \nonumber\\
&&= \int_{t_i}^{t_{\star}} \left[ \Hcl_{f}(\R_{\star},\Pcl_{f}(\R_{\star}, \tau), \tau) - \Hcl_{i}(\R_{\star},\Pcl_{i}(\R_{\star}, \tau), \tau)\right] d\tau \nonumber\\
&& +S_f(\R_f, t_f; \R_{\star}, t_i)+S_i(\R_{\star}, t_i; \R_i, t_i)  . 
\end{eqnarray}
Usually in the case of multiphoton ionization, $t_{\star}$ is complex and $\R_{\star}$ is real. Therefore, the last two terms affect only the phase and does not contribute to the probability.

Finally, since $\Hcl_f$ and $\Hcl_i$ are  the QC limits of $E_f$ and $E_i$ (this will be demonstrated below), we conclude that the exponential factors of Eqs. (\ref{AmplitudeFinalExpr}) and (\ref{ImTime}) indeed coincide within the QC approximation.

The wave function 
\begin{eqnarray}\label{PsiQC}
\Psi_{qc}(\R, t) = \int  F_i(\R, t; \R', t_i) e^{\frac{i}{\hbar}S_i(\R, t; \R', t_i)} \phi_{in}(\R') d^3\R',
\end{eqnarray}
is the (leading-order term) QC solution of Eq. (\ref{FullSchEq}) with the initial condition $\Psi_{qc}(\R, t_i) = \phi_{in}(\R)$. Employing Eq. (\ref{HJEq}) and  bearing in mind that $\Hcl_{i}=\Hcl_{i}(\R,\Pcl_{i}, t)$ does not depend on $\R'$, we obtain
\begin{eqnarray}\label{WKBAdiabaticTermState}
\h \Psi_{qc}  = i\hbar \partial_t \Psi_{qc}\left[ 1 + O(\hbar)\right] = \Hcl_{i} \Psi_{qc} \left[ 1 + O(\hbar)\right].
\end{eqnarray} 
Since we have freedom of choosing the initial condition $\phi_{in}(\R)$, there are in general infinitely many wave functions [Eq. (\ref{PsiQC})] that satisfy Eq. (\ref{WKBAdiabaticTermState}). Comparing Eqs. (\ref{WKBAdiabaticTermState}) and (\ref{AdiabaticTermStateFullH}), and  taking into account the latter, we formulate the following property of the adiabatic term and state of a given quantum system {\it within the QC limit: there exists only one adiabatic term, which is equal to the classical Hamiltonian, and any solution of the corresponding Schr\"{o}dinger equation is also an adiabatic state that corresponds to this adiabatic term} (i.e., the adiabatic term is infinitely degenerate). Note that this property is completely ruled out once the general form of the QC propagator (\ref{GenaralQCPropagator}) is considered.

The property stated above, nevertheless, merely accentuates the fundamental difference between the QC and adiabatic approximations.   As mentioned in the introduction, the adiabatic approximation allows us to obtain the solution of the non-stationary Schr\"{o}dinger equation as an  asymptotic series in terms of the small parameter $\omega/\omega_{at}$; however, the QC approximation is a method of obtaining an asymptotic expansion of the solution with respect to the small parameter $\hbar$. These two series are dissimilar in the general case. 


In conclusion, we clarify a connection between Eq. (\ref{AmplitudeFinalExpr}) and previous studies. Let $\h$ be the Hamiltonian of an atom in a low-frequency laser field,
\begin{equation}\label{HAtominField}
 \h(\varphi) = \hat{\P}^2 /2 -  Z/r + \hat{V}_L(\varphi),
\end{equation}
where the term $\hat{V}_L(\varphi)$ describes the interaction between the laser field and the atom. In Ref. \cite{Delone_1985}, the Keldysh theory \cite{Keldysh_1965} has been formulated within the Dykhne approach by choosing $E_f(\varphi) = \A^2(\varphi)/2$ as the energy of classical oscillations of a free electron in the laser field and $E_i(\varphi)=-I_p$, 
where 
$
\A(\varphi) = - ({\bf F}/\omega)\sin\varphi
$
is the vector potential of the linearly polarized laser field, $F$ is the strength of the laser field, and $I_p$ is the ionization potential. To obtain a general expression for single-electron spectra (within exponential accuracy), this formulation has been generalized in Ref. \cite{Bondar2008} by setting $E_f(\varphi) = \left[ \K + \A(\varphi) \right]^2/2$ and $E_i(\varphi)=-I_p$, where $\K$ being the momentum of the electron at the detector. Equation (\ref{AmplitudeFinalExpr}) rigorously justifies such employment of the Dykhne  theory. Indeed, partitioning Hamiltonian (\ref{HAtominField}) such that $\hat{V} = -Z/r$ [see Eq. (\ref{Partition})], we conclude that according to Eq. (\ref{H0AdiabaticState}), $\ket{\phi_f(\varphi)}$ is the Volkov wave function; thus, $E_f(\varphi) = \left[ \K + \A(\varphi) \right]^2/2$ is an exact equality within the given partitioning of the Hamiltonian. However, $E_i(\varphi)=-I_p$ can be obtained only when the dynamical Stark shift is ignored. 

Let the Coulomb potential energy be represented as
\begin{equation}\label{PopovPartitioning}
-Z/r = V_{shr}(r) + V_{lng}(r),
\end{equation}
where the potential $V_{lng}(r)$ has a long-range behavior identical to $-Z/r$, but no 
singularity at the origin, and $V_{shr}(r)$ is a singular but short-range potential. Partitioning [Eq. (\ref{PopovPartitioning})] has been originally introduced in Ref. \cite{Popov1968}; recently, the similar partitioning has been exploited in the case of two-electron systems \cite{Smirnova2008a}. For the following representation of Hamiltonian (\ref{Partition}):
$$
\h_0(\varphi) = \hat{\P}^2/2 + V_{lng}(r) + \hat{V}_L(\varphi), \quad \hat{V}(\varphi) = V_{shr}(r),
$$
Equation (\ref{AmplitudeFinalExpr}) is an adiabatic formulation of Eq. (23) of Ref.  \cite{Popov1968}. The last equation represents the amplitude of single-electron ionization that accounts for the Coulomb corrections. Therefore, Eq. (\ref{AmplitudeFinalExpr}) allows for further investigations ({\it beyond the QC approximation}) of the important problem of the influence of a Coulomb potential on ionization rates \cite{Popruzhenko2008a, Smirnova2008a, Popov2005, Popov2004}.


We are indebted to M. Yu. Ivanov and S. V. Popruzhenko  for  highly stimulating discussions. Financial support to D.I.B. by an NSERC SRO grant is gratefully acknowledged. 


\end{document}